\documentclass[twocolumn]{aa}

\usepackage{amssymb}
\usepackage{natbib}
\bibpunct{(}{)}{;}{a}{}{,} 
\usepackage{graphicx}

\newif \ifAMStwofonts \AMStwofontstrue

\begin{document}

\title{Modeling time delays in the X-ray spectrum of\\
       the Seyfert galaxy MCG-6-30-15}

\author{R.~W. Goosmann,\inst{1,2} B. Czerny,\inst{3} V. Karas,\inst{1}
        \and G. Ponti,\inst{4,5}}

\institute{
  $^1$~Astronomical Institute of the Academy of Sciences,
       Bo\v{c}n\'{\i}~II~1401, CZ--14131 Prague, Czech Republic\\
  $^2$~Observatoire de Paris, Section de Meudon, LUTH, 5 place Jules Janssen,
       F--92195 Meudon Cedex, France\\
  $^3$~Copernicus Astronomical Center, Bartycka 18, P--00--716 Warsaw, Poland\\
  $^4$~Dipartimento di Astronomia, Universit\`a di Bologna, Via Ranzani 1,
       I--40127 Bologna, Italy\\
  $^5$~INAF--IASF Bologna, via Gobetti 101, I--40129, Bologna, Italy\\
}

\authorrunning{Goosmann et al.}

\titlerunning{Modeling X-ray time delays in MCG-6-30-15}

\date{Received ... / Accepted ... }

\abstract{}
{We propose a reflection model of the time delays detected during an
  exceptionally bright, single flare in MCG-6-30-15. We consider a scenario in
  which the delays of the hard X-rays with respect to the soft X-rays are
  caused by the presence of the delayed reflection component.}
{We employ a model of the flare, which is accompanied by reprocessed
  emission. We consider two geometries/thermal states of the reprocessing
  medium: a partially ionized accretion disk surface and a distribution of
  magnetically confined, cold blobs.}
{The reprocessing by cold blobs predicts positive time delays and a
  saturation in the time delay -- energy relation, which is likely present in
  the data. The model requires a strong reflection component and relies on the
  apparent pivoting of the combined primary and reflected spectrum. The
  reflection by the ionized disk surface does not reproduce the observed
  delays. We discuss the relation between the two reflection scenarios and
  argue that they are both present in MCG-6-30-15.}
{}


\maketitle


\section{Introduction}

It has been known for a long time that Active Galactic Nuclei (AGN) show
strong non-periodic variability in the X-ray band on timescales ranging from
years down to hours and minutes (e.g. Lawrence et al. 1987; McHardy \& Czerny
1987; Mushotzky et al. 1993; Vaughan et al. 2003a; Kaastra et al. 2004).   

The predominant power-law character of the X-ray emission of most Seyfert
galaxies and quasars can be successfully modeled assuming Compton
scattering of soft photons by a hot plasma (e.g. Shapiro, Lightman \&
Eardley 1976; Sunyaev \& Titarchuk 1980). In the accretion disk scenario of
AGN the soft photons are naturally provided by the disk emission whilst the
plasma can be identified with a hot corona of a temperature $\sim 10^9$~K. The
detection of the high energy cut-off in a number of radio quiet AGN, with
values between 60 and 300~keV (e.g. Zdziarski et al. 1995; Risaliti 2002;
Deluit \& Courvoisier 2003; Soldi et al. 2005; Beckmann et al. 2005), supports
the thermal origin of the X-ray emission in these sources.

The power-law shape of the X-ray spectra is a consequence of successive
Compton up-scatterings. This should lead to systematic time delays between the
hard and the soft X-ray variations as the hard X-ray photons must undergo more
scattering events. A search for such delays was performed for MCG-6-30-15 and
other Seyfert galaxies but the results are not unique.

Reliable measurements of X-ray time delays in AGN spectra require sufficiently
long observation times. For a 95 ksec observation of MCG-6-30-15 with {\it
  XMM-Newton} the quality of the data did not allow to confirm any spectral
delays (Ponti et al. 2004). They were only reported after a follow-up 300 ksec
observation when Vaughan et al. (2003b) detected delays by 200 sec on
timescales of $10^4$~sec. {\it Frequency-dependent} time delays were found in
a number of sources (NGC~7469, Papadakis, Nandra \& Kazanas 2001; NGC~4051,
McHardy et al. 2004; MCG-6-30-15, Vaughan et al. 2003b; NGC~3783, Markowitz
2005; Ark~564, Ar\'evalo et al. 2006), for a frequency range of $10^{-6}$ to
$10^{-3}$ Hz. These delays turn out to be roughly proportional to the Fourier
period and they increase with the energy span considered (max. between 0.2 to
12~keV) up to a few thousand seconds.

The dependence of the delay time on the Fourier frequency cannot be
accommodated within the frame of a simple Comptonization model that assumes
electron clouds of a uniform density and temperature. In such models the
observed hard-to-soft time delay is independent of the Fourier frequency
(Miyamoto et al. 1988). Therefore, two types of more sophisticated models were
proposed:

\begin{itemize}

  \item[\textbullet] an extended Comptonizing region with the appropriate
  density and temperature stratification (extended spherical region, Kazanas
  et al. 1997; advection-dominated inflow-outflow model, Blandford \& Begelman
  1999; extended coronal emission model, Arevalo \& Uttley 2006)\\

  \item[\textbullet] models of a local spectral evolution due to coherent
  changes in a single compact emission region (single coronal flare evolution,
  Poutanen \& Fabian 1999, \.Zycki 2003; inflow of a single blob, B\"ottcher
  \& Liang 1999, \.Zycki 2004)

\end{itemize}

The local models predict a pivoting of the hard X-ray spectrum, and the
pivoting around the appropriate frequency can explain the apparent time delays,
as argued by K\"ording \& Falcke (2004).

The detection of a single gigantic flare in the lightcurve of MCG-6-30-15
(Ponti et al. 2004) offers a specific insight into the delay problem assuming
that in this exceptional case we see a localized flare event (instead of a
random peak in the net flux of large flare distributions). This assumption
is inspired by the rather symmetric shape of the lightcurve during the flare
event and by its short time scale of only $\sim 2200$~sec. Ponti et al. (2004)
detected spectral time delays during the flare period and found that the hard
X-rays were lagging the soft band. They interpret this as a result of multiple
photon scattering within the flare itself. 

Using the same data, we discuss in this paper an alternative possibility to
explain the time delays. While the models listed above focus primarily on the
Compton up-scattering in the hot plasma, we investigate a possible role of the
reprocessed component for the observed time delays. Unlike the primary, the
reprocessed radiation arises in a relatively cooler medium and thus has a
different spectral shape. The spatial distance between the primary source and
the reprocessing site causes a delay between the two components and a time
evolution of the combined (i.e. primary plus reflection) spectrum. We will
show that this leads to an apparent effect of energy-dependent time delays. We
investigate two types of reprocessing: a strongly ionized reflector and a
weakly ionized, cold reflector. 

After recalling the analysis of the data (Sect.~2) we describe a simple model
(Sect.~3) to find general trends of the spectral time delays between various
energy bands. We discuss their dependence on details of the model setup for
the two reprocessing scenarios (Sect.~4). We then show how the spectral delays
of the observed flare can be reproduced by our model (Sect.~5) and we discuss
and summarize our results (Sect.~6).

\section{Observation and data reduction}

The data were collected during the observation performed on June
11$^{\rm th}$ 2000, with the EPIC MOS1, MOS2 and pn cameras of {\it XMM-Newton}
(Str\"uder et al. 2001) operated in timing, full-frame, and small window
modes. The source was observed for 95~ksec and the spectral data were first
published in Wilms et al. (2001) whilst Ponti et al. (2004) performed a timing
analysis.

The data were re-reduced and screened using the SAS software v6.5.0. We only
used the pn data as there was no significant pile--up present. The single and
double pixel events (i.e. pattern $\leq$4) were selected. Source and
background counts were extracted from two circular regions of 45$\arcsec$
radius. The background events were taken from a region far from the source but
within the same chip.

\begin{figure}
  \hskip +0.12cm
  \includegraphics[width=6cm,angle=-90]{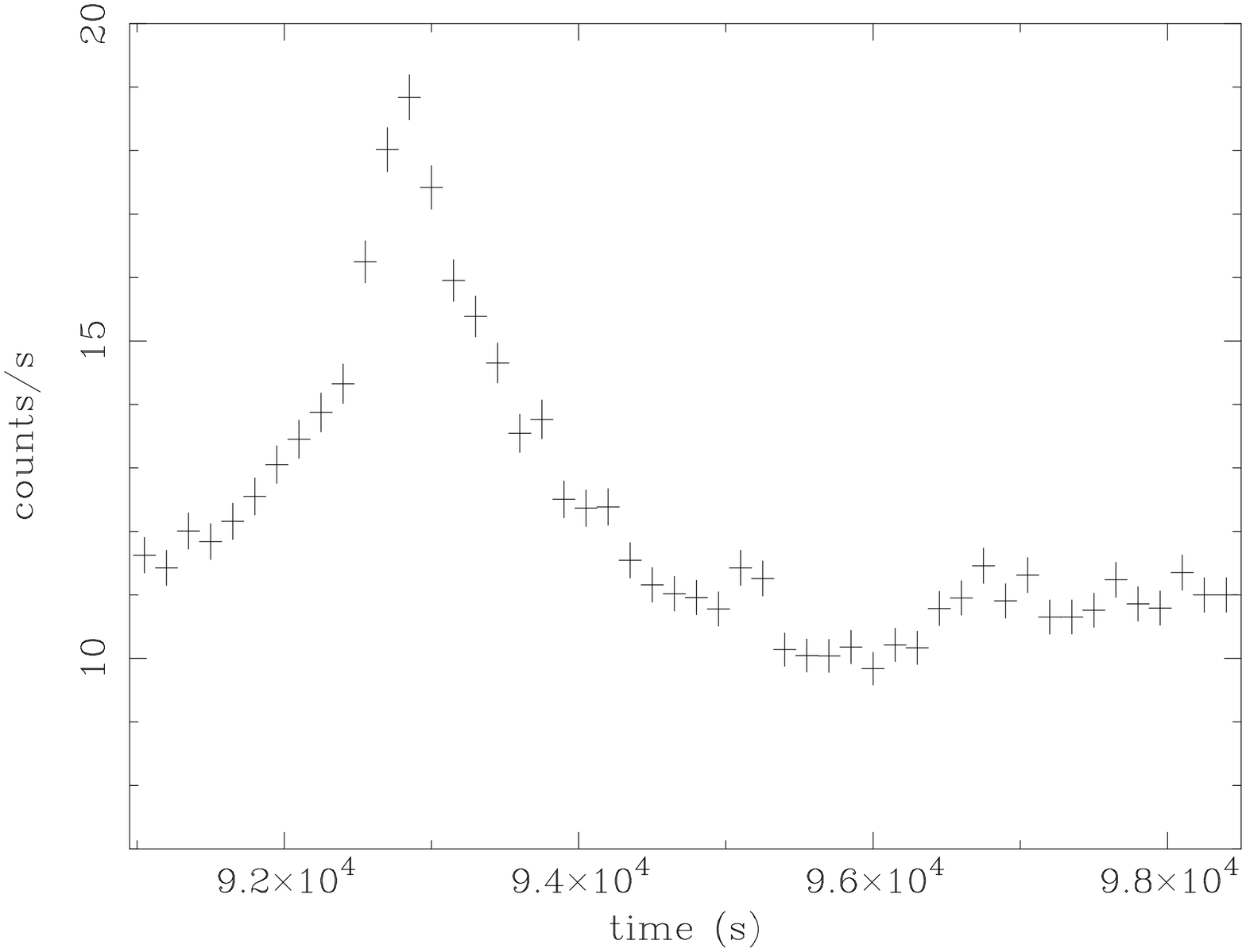}
  \vskip 0.3cm
  \includegraphics[width=5.2cm,angle=-90]{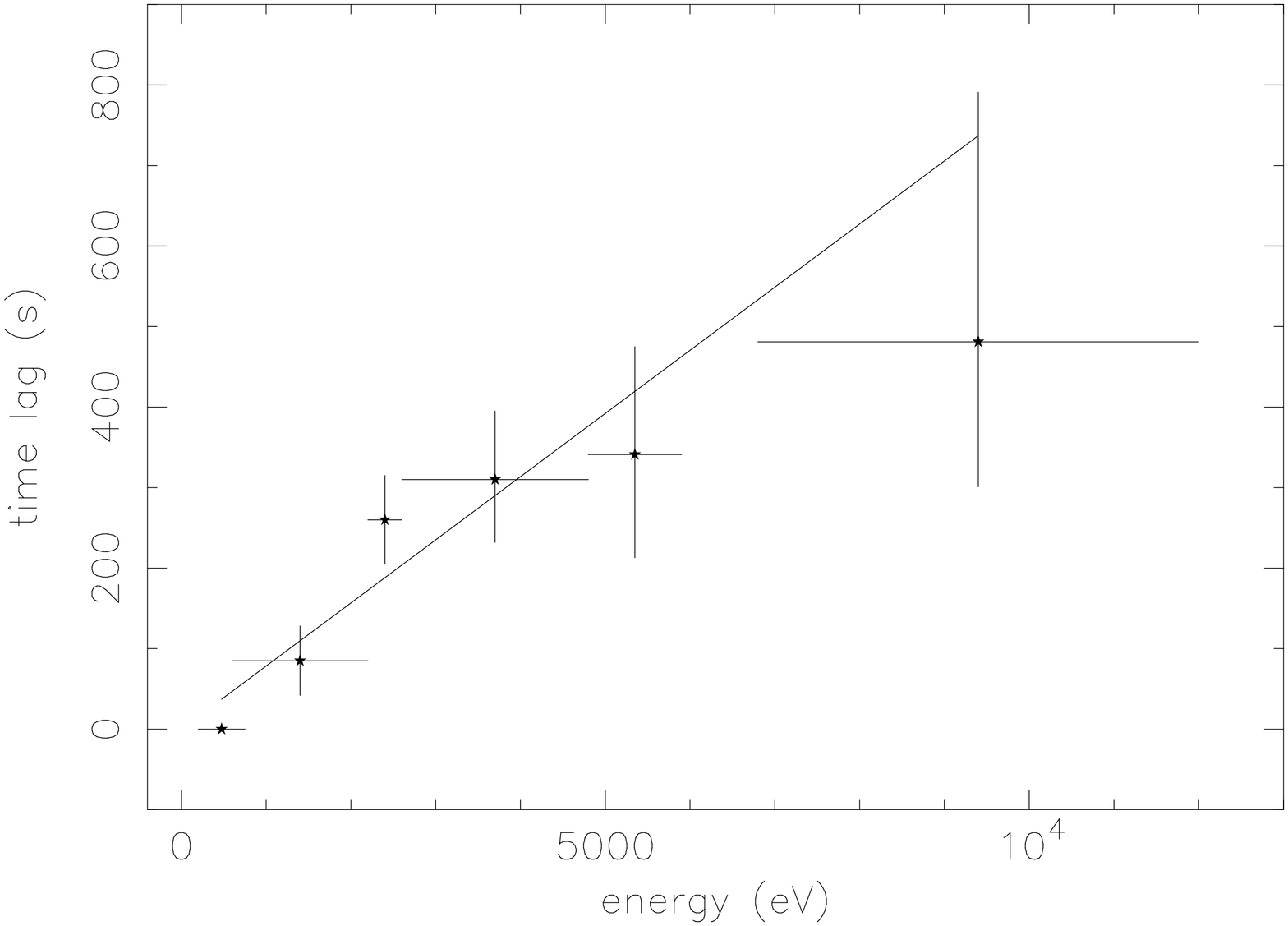}
  \vskip 0.5cm
  \caption{Top: the light-curve over 0.2--10~keV during the period of the
  strong flare in MCG-6-30-15. Bottom: energy-dependent time delays computed
  by \citet{ponti2004} together with a linear fit representing Comptonization
  inside the flare. \label{fig:mcg-flare}}
\end{figure}

During the observation MCG-6-30-15 showed a bright flare lasting for $\sim
2000$~sec. The lightcurve over the flare period is shown in
Fig.~\ref{fig:mcg-flare} (top). It was decomposed into six lightcurves in
separate energy bands, and X-ray time delays between them were calculated from
cross-correlation functions (Ponti et al. 2004). The delays increase with
difference in energy (Fig.~\ref{fig:mcg-flare}, bottom). The photons at 10~keV
lag by a few hundred seconds with respect to those below 0.57~keV.

\section{The model}

\label{sect:model}

We assume that the flare is created in a compact region at some distance from
the reprocessing matter. The resulting primary component $I_{\rm{}p}$ partly
shines directly toward a distant observer and partly gives rise to the
Compton reflection/reprocessed component $I_{\rm{}r}$. The response of
$I_{\rm{}r}$ to spectral variations of $I_{\rm{}p}$ is delayed by the
light travel time between the flare source, the reprocessing sites, and the
observer. This time delay depends on the geometry in the vicinity of the flare
and also on the viewing angle. An extended reprocessing region should lead to
some level of smearing of the original signal, i.e. the duration of the
brightening of the reflected component will be longer than the duration of the
flare itself.

We introduce a simple mathematical parameterization of $I_{\rm{}p}$ and
$I_{\rm{}r}$ as a function of energy $E$ and time $t$. For both components we
assume a power-law dependence in energy with indexes $\alpha_{\rm{}p}$ and
$\alpha_{\rm{}r}$. For $I_{\rm{}r}$ a broken power-law is also possible. It
involves two branches with indexes $\alpha_{\rm{r}_1}$ and $\alpha_{\rm{r}_2}$
that connect at the break energy $E_{\rm br}$. This parameterization neglects
particular reprocessing features, such as the emission lines, but represents
rather well the overall shape of the continuum for a neutral or partially
ionized reflector. 

We imagine the time modulation of $I_{\rm{}p}$ and $I_{\rm{}r}$ as consecutive
events. They are modeled by Lorentzians, $\mathcal{L}_{\rm{}p}(t)$ and
$\mathcal{L}_{\rm{}r}(t)$, which take the half-width parameter $T_{\rm f}$ and
the time $t_0$ marking the maximum of the primary emission. The primary leads
the reprocessed spectrum by a delay $\delta$ giving a measure of the average
light travel time between the flare source and the reprocessing sites. As
there are many possible light paths for the reprocessing the width of
$\mathcal{L}_{\rm{}r}$ can be broadened by a factor $b$. Note that this
broadening is also influenced by the viewing angle of the distant
observer. The equations for $I_{\rm{}p}$ and $I_{\rm{}r}$ read as follows:

\begin{eqnarray}
  I_{\rm{}p}(E,t) & = & \mathcal{L}_{\rm{}p}(t) E^{-\alpha_{\rm{}p}},\\
  I_{\rm{}r}(E,t) & = & N \mathcal{L}_{\rm{}r}(t) \times
  \left \{
  \begin{array}{ll} 
    E^{-\alpha_{\rm{r}_1}}  & {\rm for} \; E \leq E_{\rm br},\\
    E_{\rm br}^{\alpha_{\rm{r}_2}-\alpha_{\rm{r}_1}}
    E^{-\alpha_{\rm{r}_2}}  & {\rm for} \; E > E_{\rm br},
  \end{array}
  \right. \nonumber
\end{eqnarray}

\noindent where

\begin{eqnarray}
  \mathcal{L}_{\rm{}p}(t) & = & 
  \frac{T_{\rm f}^2}{(t-t_0)^2+T_{\rm f}^2},\\
  \mathcal{L}_{\rm{}r}(t) & = &
  \frac{b^2 T_{\rm f}^2}{(t-t_0-\delta)^2+b^2 T_{\rm f}^2}.
\end{eqnarray}

\begin{table}
  \caption{Energy bands and fixed parameters of the model
           \label{tab:constants}}
  {\small
  \centering
  \begin{tabular}{cc|cc}
    \hline
    \hline
      Energy band & Range [keV] & Fixed parameter & Value  \\
    \hline
      $\Delta E_1$ & 0.2 -- 0.57 & $E_{\rm{}min}$ & 0.2~keV\\
      $\Delta E_2$ & 0.6 -- 2.2  & $E_{\rm{}max}$ & 12~keV\\
      $\Delta E_3$ & 2.2 -- 2.6  & $\alpha_{\rm{}p}$ & 0.9, 1.3\\
      $\Delta E_4$ & 2.6 -- 4.8  & $t_0$ & 2200~sec\\
      $\Delta E_5$ & 4.8 -- 6.8  & $T_{\rm f}$ & 500~sec\\
      $\Delta E_6$ & 6.8 -- 12.0 & ~ &~ \\
    \hline
  \end{tabular}
  }
\end{table}

A factor $N$ is included in $I_{\rm{}r}$ to normalize the reflected component
against the primary. We define $N$ with respect to the ratio, $K$, between
$I_{\rm{}p}$ and $I_{\rm{}r}$ integrated over time and energy: 

\begin{eqnarray}
  K & = &
    \frac{ N b \left[
    \int\limits_{E_{\rm{}min}}^{E_{\rm{}br}} dE' E'^{-\alpha_{\rm{}r_1}} +
    \int\limits_{E_{\rm{}br}}^{E_{\rm{}max}} dE'' E''^{-\alpha_{\rm{}r_2}}
    \right]
    }
    {
    \int\limits_{E_{\rm{}min}}^{E_{\rm{}max}} dE \, E^{-\alpha_{\rm{}p}}}.
   \label{eqn:K-value}
\end{eqnarray}

\noindent with $b = \left[ \int \mathcal{L}_{\rm{}r}(t') dt' \right] / \left[
\int \mathcal{L}_{\rm{}p}(t) dt \right]$. For a given value of $K$, the
normalization $N$ is a function of the energy limits, the broadening, and the
spectral indexes. A distant observer detects the sum

\begin{equation}
  I_{\rm{}obs}(E,t) = I_{\rm{}p}(E,t) + I_{\rm{}r}(E,t).
\end{equation}

We specify the energy bands, $\Delta E_{\rm i}$, with $i = 1,...,6$, and
construct light-curves

\begin{equation}
  L_{\rm i}(t) = \int \limits_{\Delta E_{\rm i}} I_{\rm{}obs}(E,t) dE.
\end{equation}

To investigate the time delay between the spectral response in different
energy bands we compute cross-correlation functions

\begin{equation}
  F_{\rm{}CCF}^{\rm ij}(\tau) = \frac{\int
  \limits_{-\tau_{\rm{}max}}^{+\tau_{\rm{}max}} L_{\rm i}(t) L_{\rm j}
  (t-\tau) dt} {\sqrt{\int \limits_{-\tau_{\rm{}max}}^{+\tau_{\rm{}max}}
  L_{\rm i}^2(t')\;dt'} \times \sqrt{\int
  \limits_{-\tau_{\rm{}max}}^{+\tau_{\rm{}max}} L_{\rm j}^2(t'')\;dt''}},
\end{equation}

\noindent where the denominator is a normalization factor and $\tau_{\rm{}max}$
must be large enough for $F_{\rm{}CCF}^{\rm ij}(\tau)$ to converge. The global
maximum of $F_{\rm{}CCF}^{\rm ij}(\tau)$ marks the time delay between the
energy bands $\Delta E_{\rm i}$ and $\Delta E_{\rm j}$.

\section{Results}
\label{sec:TL-results}

Our model contains various parameters. Some of them are fixed by the flare
observation described in \citet{ponti2004}. The total energy range, and the
energy channels are specified as in the analysis: $E_{\rm{}min} = 0.2$~keV,
$E_{\rm{}max} = 12$~keV and the energy bands are listed in
Table~\ref{tab:constants}. We adopt the time of the flare maximum at $t_0 =
2200$~sec and the half width $T_{\rm f} = 500$~sec. The slope of the primary
emission is not known very accurately due to the complex X-ray spectrum of
MCG-6-30-15. It correlates with the hard X-ray flux, and rises up to $\sim
1.3\,$ \citep{vaughan2001,shih2002}. We thus choose $\alpha_{\rm{}p} = 1.3$ as
the upper limit of the slope. As the lower limit we set $\alpha_{\rm{}p} =
0.9$, which is consistent with {\it Beppo-SAX} data suggesting slopes $\sim
0.94$ \citep{fabian2002a}, and also with the average slope in a large sample
of radio-quiet AGN \citep{nandra1997,george2000,piconcelli2005}. The fixed
parameters of the model are summarized in Table~\ref{tab:constants}.

\subsection{Delays for a highly ionized reflector}
\label{sec:ion-disk}

\begin{figure}
  \centering
  \includegraphics[width=8cm]{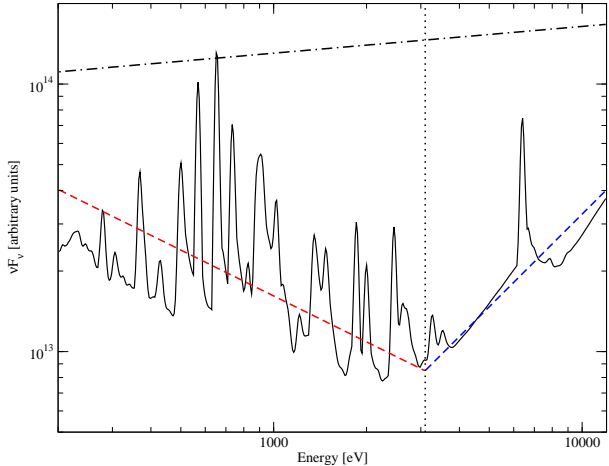}
  \caption{An example of ionized reflection: the primary $F_{\rm p}(E) \propto
  E^{-0.9}$ (dot-dashed) and the local reprocessing (solid) of the ionized
  reflector scenario. The dashed line denotes the power-law fit $F_{\rm r}(E)
  \propto E^{-\alpha_{\rm r}}$ adopted for the delay modeling. Below 3.1~keV
  the slope $\alpha_{\rm r_1} = 1.57$ and above $\alpha_{\rm r_2} =
  -0.15$. \label{fig:locspec}}
\end{figure}

Highly ionized reflection is likely to form when a strong flare is localized
not far from the accretion disk. An example of the reflected spectrum for
such a case is shown in Fig.~\ref{fig:locspec}. The spectrum is calculated
using the codes {\sc Titan} and {\sc Noar} \citep{dumont2000,dumont2003}. The
disk vertical structure is determined with an improved version of the code
given in \citet{rozanska2002}. We assume $M = 10^7 M_\odot$
\citep{mchardy2005} and $a/M = 0.998$ \citep{fabian2002a,miniutti2003,
fabian2003}. The flare is located at the disk radius $r = 18 \, R_{\rm g}$. 

The reflected component has a typical spectral shape for reprocessing by
highly ionized material: it is relatively hard above $\sim 3 $~keV and steep
(soft) at lower energies:

\begin{equation}
  I_{\rm{}r}(E,t) = N \mathcal{L}_{\rm{}r}(t) \times 
  \left \{ 
  \begin{array}{ll} 
    E^{-1.57}  & {\rm for} \; E \leq 3.1 \; {\rm keV},\\
    E^{+0.15}  & {\rm for} \; E > 3.1 \; {\rm keV}.
  \end{array}
  \right. \nonumber
\end{equation}

The relative normalization of the primary and the reflected component 
is taken from the modeling shown in Fig.~\ref{fig:locspec}. We obtain $K \sim
0.15$ if the flare radiates isotropically. With $b = 2$ this leads to a
normalization of $N \sim 0.022$ ($N \sim 0.025$) for $\alpha_{\rm{}p} = 0.9$
($\alpha_{\rm{}p} = 1.3$). However, large values of the equivalent width of
the iron line (Wilms et al. 2001, Reynolds et al. 2004) indicate that the
reflection may be strongly enhanced with respect to isotropic emission
(e.g. Fabian et al. 2002a, Ballantyne et al. 2003, Taylor, Uttley \& McHardy
2003). Therefore, we keep $N$ as a free parameter of the model.

\begin{figure*}
  \centering
  \includegraphics[height=13.3cm]{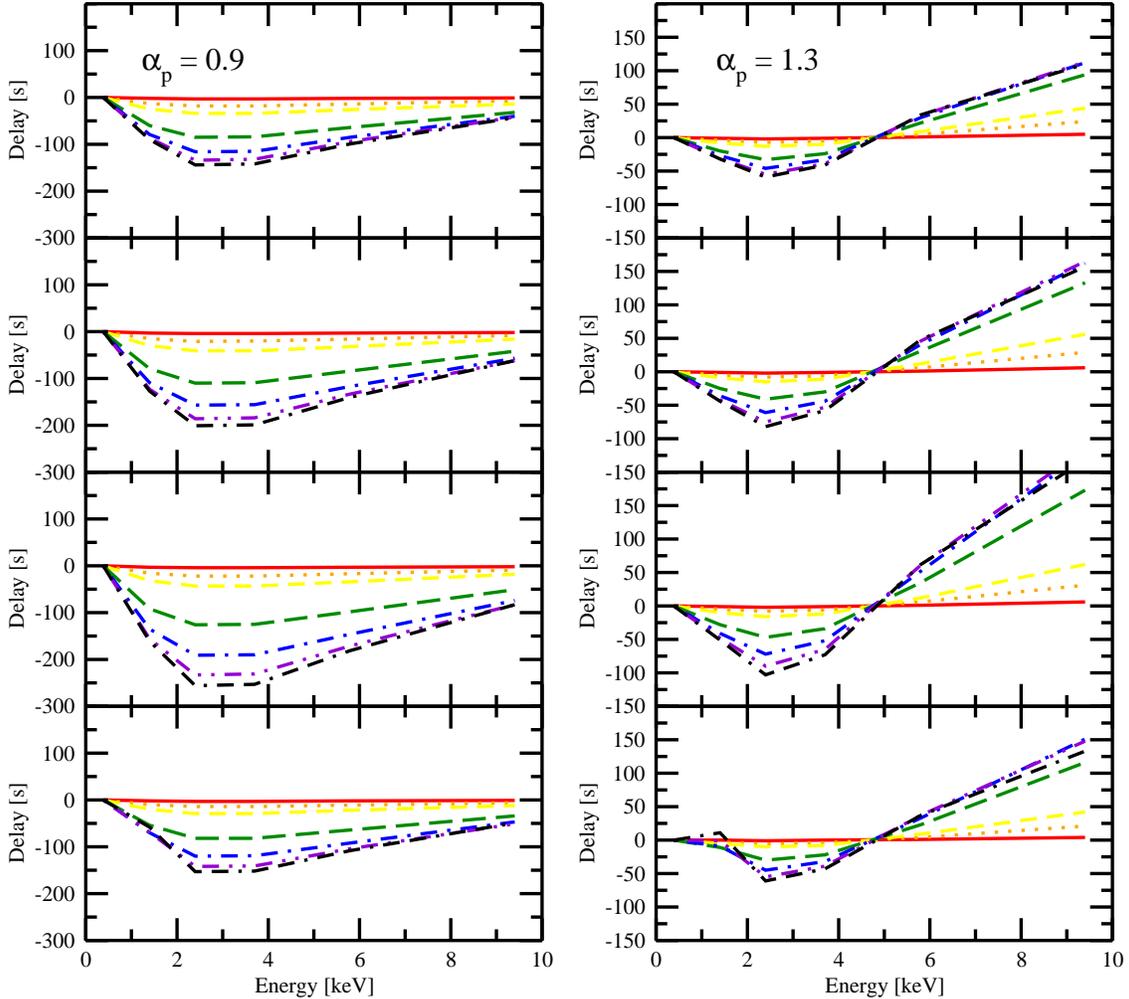}
  \caption{Energy-dependent time delays for a highly ionized reflector with
  two slopes of the primary spectrum. Left: $\alpha_{\rm{}p} = 0.9$, right:
  $\alpha_{\rm{}p} = 1.3$. From top to bottom the panels represent intrinsic
  delays $\delta = $ 500~sec, 700~sec, 1000~sec, and 1700~sec. The different
  curves denote the following values for $N$: 0.01 (red, solid), 0.05 (orange,
  dots), 0.1 (yellow, short dashes), 0.3 (green, long dashes), 0.5 (blue,
  dashes/dots), 0.7 (pink, dashes/double-dots), and 0.9 (black,
  double-dashes/dots). Note that the vertical scale is different for both
  columns. \label{fig:PL-res-delay}}
\end{figure*}  

We vary $N$ between 0.01 and 0.9 and set the intrinsic delay $\delta$ to
500~sec, 700~sec, 1000~s and 1700~sec. The spectral time delays at all energy
bands $\Delta E_{\rm i}$ are obtained and normalized so that they are measured
with respect to the one at $\Delta E_1$. The same normalization was applied in
the data analysis \citep{ponti2004}. The results for a broadening of $b = 2$
and for the two values of $\alpha_{\rm p}$ are shown in
Fig.~\ref{fig:PL-res-delay}. 

Obviously, for the very low normalization values of $N = 0.022$ or $N = 0.025$
\emph{no significant time delays appear}. Noticeable delays only exist for $N$
being higher by at least one order of magnitude. But even at higher $N$ these
delay curves are very different from the ones obtained by
\citet{ponti2004}. The model mostly produces negative delays, while the
observed ones are always positive.

The behavior of the delay curves as a function of $N$ and $\delta$ is partly
analogous in both cases of $\alpha_{\rm p}$. The overall normalization of the
curves increases with $\delta$ until a maximum at $\sim 1000$~sec is reached,
then, for even higher $\delta$, the normalization decreases again. It also
rises with $N$ but saturates when $N$ goes beyond unity.

There are also some important differences between two cases of $\alpha_{\rm
p}$. For $\alpha_{\rm p} = 0.9$ the delays are negative at all energies and the
shape of the curves is concave over the whole energy range. The minimum delay
occurs around 3~keV. In the case of $\alpha_{\rm p} = 1.3$ the shortest delays
appear at a slightly lower value. The curves rise more quickly at intermediate
energies and their curvature becomes convex around 5~keV. At high energies,
long positive delays are obtained. 

\begin{figure}[h!]
  \vskip 0.4cm
  \centering
  \includegraphics[width=7cm]{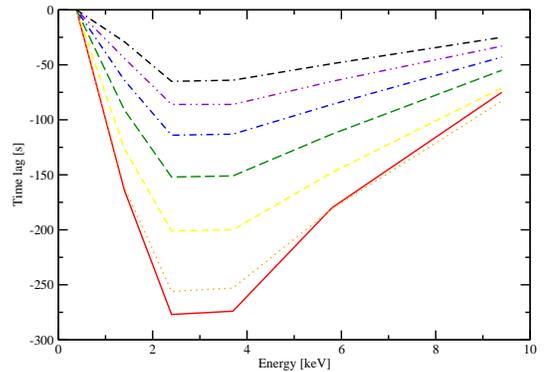}
  \caption{Time delay curves derived for various values of $b$ in the scenario
  of a highly ionized reflector. The normalization $Nb$ of the total energy
  released is kept constant at $Nb = 1.8$. The bottom curve (red, solid)
  represents the value $b = 1.5$. From bottom to top, the value increases in
  steps of 0.5 for each curve and ends up at $b = 5$ for the highest curve
  (black dots/double-dashes). \label{fig:PL-res-broaden}}
  \vskip -0.6cm
\end{figure}

We vary the broadening parameter $b$ while fixing the time-integrated energy
output of the reprocessed component. Thus, for each value of $b$ we find the
corresponding $N$ from Eqn.~(\ref{eqn:K-value}). We consider seven values for
$b$ and $N$ with $Nb = 1.8$. The resulting delay curves for $\alpha_{\rm p} =
0.9$ and $\delta = 1000$~sec are shown in Fig.~\ref{fig:PL-res-broaden}. For a
given value of $Nb$ a whole range of curves is obtained. However, the
characteristic concave curvature and the minimum around 3~keV are always
preserved. Conducting the same investigation for other values of $Nb$ delivers
similar results. Therefore, varying $b$ does not help to overcome the
discrepancy between the modeled time delays and those observed during the
flare in MCG-6-30-15.

\subsection{Delays for a weakly ionized/neutral reflector}
\label{sec:clump-disk}

In our second scenario we investigate the time delays obtained for a weakly
ionized reflector. This is a likely situation when the flare is located
far from the accretion disk and/or the cold plasma is magnetically confined
and denser than the plasma at the surface of an irradiated standard disk.

\begin{figure*}
  \centering
  \includegraphics[width=13.3cm]{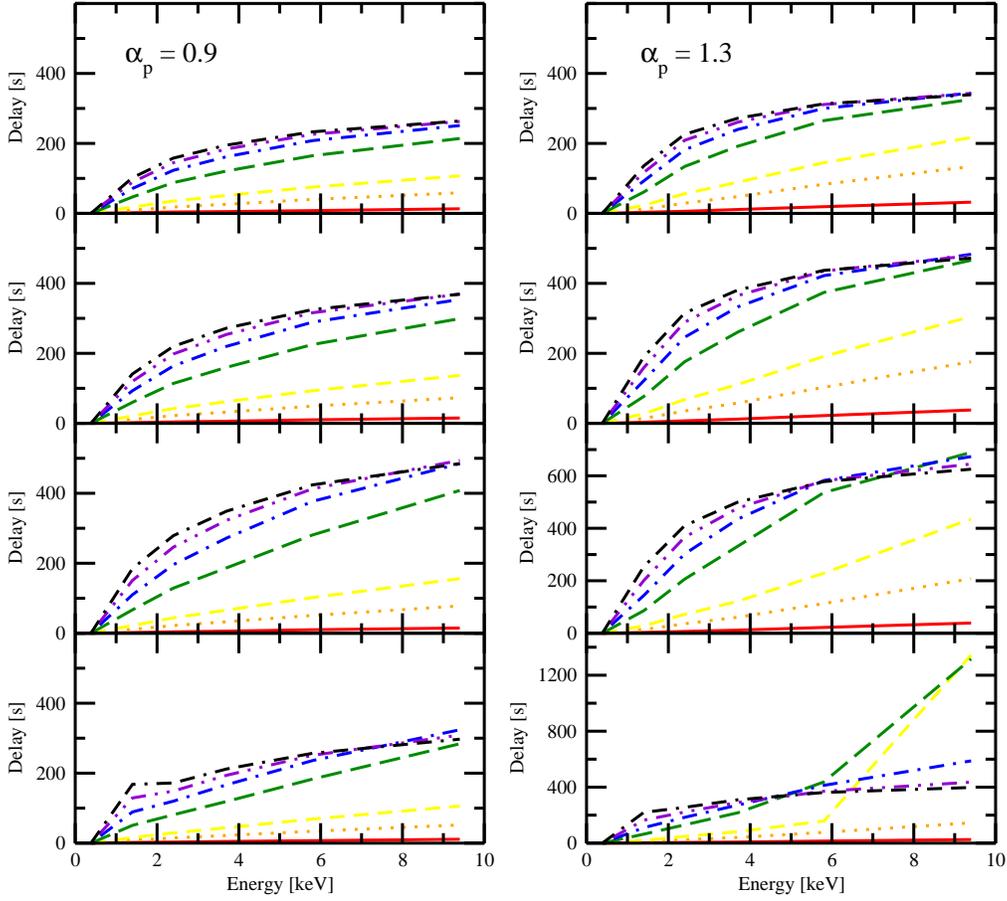}
  \caption{Energy-dependent time delays for a weakly ionized/neutral reflector
  for two slopes of the primary spectrum. Left: $\alpha_{\rm{}p} = 0.9$,
  right: $\alpha_{\rm{}p} = 1.3$. From top to bottom the panels represent
  intrinsic delays $\delta = $ 500~sec, 700~sec, 1000~sec, and 1700~sec. The
  different curves denote the following values for $N$: 0.01 (red, solid),
  0.05 (orange, dots), 0.1 (yellow, short dashes), 0.3 (green, long dashes),
  0.5 (blue, dashes/dots), 0.7 (pink, dashes/double-dots), and 0.9 (black,
  double-dashes/dots). Note the change in vertical scale for the last two
  panels of the right column. \label{fig:res-delay}}
\end{figure*}

As a specific example of the reprocessed component we use the model
of irradiated magnetically confined, cold clumps. The radiative transfer
calculations for such a medium were performed by \citet{kuncic1997}.
An important effect in their modeling is that successive reprocessing can
significantly lower the spectral energy in the soft X-ray range leading to
very hard spectral slopes of $I_{\rm r}$. This hardening depends on the number
density of the clouds in the medium. Here we adopt the parameterization

\begin{equation}
  I_{\rm r} = N \mathcal{L}_{\rm{}r}(t) E^{-0.1},
\end{equation}

\noindent which is inferred from the case of 50 individual clouds on the line
of sight. The index $\alpha_{\rm r} = 0.1$ is taken from fitting the
corresponding curve in Fig.~2 of \citet{kuncic1997} between 0.2~keV and 12~keV
with a power-law. We compute delay curves for $\delta$ set to 500~sec,
700~sec, 1000~sec, and 1700~sec. The values of the normalization $N$ are taken
between 0.01 and 0.9 and the broadening parameter is set to $b = 2$. The
results are shown in Fig.~\ref{fig:res-delay}.

As for the highly ionized reflector the very low values of $N = 0.022$ and $N =
0.025$ lead to \emph{no significant time delays}. But for higher $N$ the curves
look very different from the previous scenario. They resemble a lot more the
observed delays found by \citet{ponti2004}. Note that strong reflection was
also required to model the energy-dependent fractional variability amplitude
in MCG-6-30-15 (Goosmann et al. 2006), and it is consistent with the large
value of the iron line equivalent width measured by several authors
(e.g. Wilms et al. 2001, Reynolds et al. 2004). Therefore, the high value of
$N$ may be justified.

The following trends can be derived from the model: the overall normalization
of the time delay curves increases with $\delta$, goes through a maximum around
$\delta = 1000$~sec and then decreases again. All curves have positive slopes
and the spectra evolve from softer to harder X-ray energies. The shape of the
curves changes with $N$. Higher values of $N$ lead to longer delays until $N$
passes unity and the curves saturate at a level, which scales with $\delta$.
For the steeper primary spectrum with $\alpha_{\rm{}p} = 1.3$ the delay curves
show the same general features and tendencies as for $\alpha_{\rm p} =
0.9$. However, for $\alpha_{\rm{}p} = 1.3$ the overall normalization of the
time delay curves is larger and the curves are generally steeper in the soft
X-ray range.

\begin{figure}
  \vskip 0.6cm
  \centering
  \includegraphics[width=7cm]{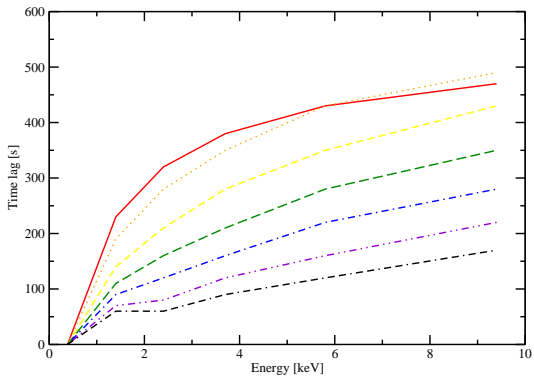}
  \caption{Time delay curves derived for various values of the broadening $b$
  in the weakly ionized/neutral reflector scenario. The normalization $Nb$ of
  the total energy released is kept constant at $Nb = 1.8$. The upper curve
  (red, solid) represents the value $b = 1.5$. From top to bottom, the value
  increases in steps of 0.5 for each curve and ends up at $b = 5$ for the
  lowest curve (black dots/double-dashes). \label{fig:res-broaden}}
\end{figure}

We investigate the influence of $b$ in a series of calculations with
$\alpha_{\rm p} = 0.9$, $Nb = 1.8$, and $\delta = 1000$~sec. The resulting time
delay curves are shown in Fig.~\ref{fig:res-broaden}. For increasing values of
$b$, the delays become smaller. This can partly be explained by the lower
normalization $N$ required to keep the same value of $Nb = 1.8$. It was shown
above that the delay curves level down for lower values of $N$. Another point
is that for larger $b$ the variations of $I_{\rm{}r}$ start earlier in
time. Thus, the variations of $I_{\rm{}p}$ are less ahead of the variations of
$I_{\rm{}r}$. As for the highly ionized reflector, conducting the same
investigation for other values of $Nb$ delivers similar results.

\section{Modeling the measured time delays of MCG-6-30-15}
\label{sec:lags-MCG}

\begin{figure}
  \vskip 0.6cm
  \centering
  \includegraphics[width=7cm]{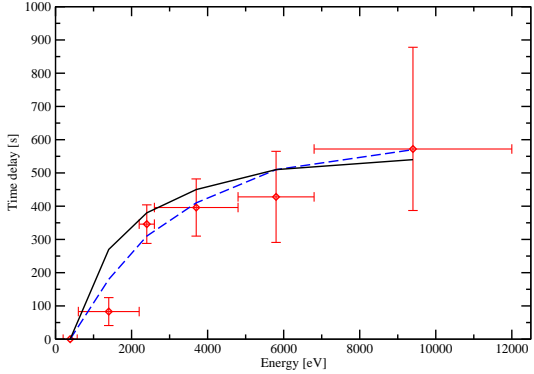}
  \caption{Satisfying representation of the measured time delays (red diamonds
  with error bars) for the flare/clumps scenario. The two curves rely on the
  following parameterization: $\delta = 1000$~sec, $b = 2.5$, and $N = 1.1$
  (solid, black) as well as $\delta = 1000$~sec, $b = 2.5$, and $N = 0.7$
  (dashed, blue). \label{fig:mcg-delay-fit}} 
\end{figure}

As shown in Fig.~\ref{fig:PL-res-delay}, the highly ionized reflector
only predicts time delays that are non-monotonic with energy. This is due to
the concave reflection spectrum of the scenario. On the other hand, a cold
reflector with a hard single power-law spectral shape gives a monotonic
increase. Certain parameter sets of the latter scenario reproduce the
observed time delays for the flare of MCG-6-30-15 quite well. In
Fig.~\ref{fig:mcg-delay-fit} we show two examples of satisfactory data
representation. The model reproduces the curvature seemingly present in the
data. This may indicate that it catches the apparent signal propagation better
than the fit by a straight line used in Ponti et al. (2004). The primary slope
of the two solutions shown was set to $\alpha_{\rm p} = 1.3$ assuming that for
the high flux state of the flare the spectrum should steepen
\citep{vaughan2001,shih2002}. 

The solution is not unique. Partly, this is due to the large error bars
attached to the data points in Fig.~\ref{fig:mcg-delay-fit}. There are three
free parameters involved: the normalization $N$, the broadening $b$, and the
intrinsic delay $\delta$. For higher accuracy of this modeling, the absolute
normalization $K$ of the reprocessed power with respect to the primary must be
determined in an independent way. Then the value of $Nb$ can be computed from
Eqn.~\ref{eqn:K-value} (which links $N$ and $b$). Obtaining $K$ for the
specific case of the observed flare in MCG-6-30-15 would require a very
detailed spectral analysis. A decomposition into primary radiation and
multiple reprocessing components using a sophisticated radiative transfer
model would be helpful. However, the flare period only lasted for $\sim
2000$~sec and the obtained spectra are not of a sufficient accuracy to allow
such an investigation.

Nevertheless, we are able to give a few robust conclusions. The model
indicates that an intrinsic time delay of 1000~sec is required. Lower
or higher values of $\delta$ do not deliver any satisfying fits. The delay of
1000~sec translates to a light travel  distance of $\sim$~60$~R_{\rm g}$,
assuming a black hole mass $M = 3.3 \times 10^6$ M$_\odot$
\citep{mchardy2005}. The broadening factor $b = 2.5$ suggests a significant
spread in the distances between the source and the reprocessor which would
agree with the multiple scattering expected in the clumpy scenario. The lower
limit of the normalization is at $N = 0.7$, hence the reprocessing has to be
strong. In the cold reflection model this implies a high covering factor of
the flare source.

\section{Discussion and summary}
\label{sec:TL-discuss}

We considered two scenarios in which the time delays of the hard to the soft
X-rays are caused by the delayed reflection component. The dependence of the
delay on energy is caused by a difference in the spectral shape between the
primary and the reprocessed spectrum. In the first scenario the reflector is
ionized, the reflection component is concave, the predicted time delays are
non-monotonic with energy, and the observed time delays are not reproduced. In
the second scenario the reflection spectrum is close to a single hard
power-law and the predicted time delays increase with energy. In this case,
the model reproduces qualitatively the observed time delay curve.

\subsection{Comptonization delay versus reprocessing delay}

Ponti et al. (2004) argued that the observed delays are caused by
Comptonization {\it inside} the flare source, whilst in our approach we
neglect the internal delay and model the effect of the delayed reflection. To
justify this method the flare has to be compact enough. The lower limit for
the flare size can be estimated from the compactness parameter \citep[see
e.g.][]{cavaliere1980, svensson1994, torricelli-ciamponi2005}: $l <
12$. During the 95 ksec {\it XMM-Newton} observation, MCG-6-30-15 was
exceptionally faint, and its 2--10~keV flux translates into a luminosity of $L
= 2.6 \times 10^{42}$ erg s$^{-1}$. The peak luminosity of the flare is
comparable to this value. Therefore, the size $D_{\rm fl}$ of the flare source
has to satisfy 

\begin{equation}
  D_{\rm{}fl} > 6 \times 10^{12} 
  \left({L \over 2.6 \times 10^{42} \rm{\ erg \ s}^{-1}} \right) 
  \left({12 \over l}\right) \quad[\mbox{cm}],
\end{equation}

\noindent so the light travel time across the flare itself can be as short as
$\sim 200$~sec and this provides a lower limit for the internal
delay. Unfortunately, we cannot estimate the number of scatterings within the
source because the observational limits for the extension of the X-ray
emission in MCG-6-30-15 are not conclusive \citep[{\it Beppo-SAX} data
indicate a cut-off energy at 100--470~keV, depending on the
model;][]{fabian2002a}.

In our successful weakly ionized/neutral reflector scenario the overall
\emph{shape} of the delay curves  is rather robust against changes of the
parameters. We find that in most cases the curve slightly flattens towards
larger energy separations. This flattening is a specific prediction of the
model and it is encouraging that it is indicated by the data. Comptonization
models on the other hand predict a linear relation in the 0.2--12~keV band,
i.e. much below the characteristic energy of the Comptonizing electrons.

\subsection{A flare located above the accretion disk}

The irradiation of a standard accretion disk by strong flare emission
was studied in numerous papers \citep[see e.g.][]{nayakshin2000a,
nayakshin2000b,ballantyne2001a,rozanska2002,collin2003}. If the flare is
located near the inner part of the disk the disk surface is strongly
ionized. Relativistic modifications of the flare spectrum and Doppler shifts 
then explain the presence of a broadened iron K$\alpha$ line
\citep[see e.g.][]{dovciak2004a,dovciak2004b}. For MCG-6-30-15 the
relativistically distorted line varies less than the continuum, which can be
accounted for by the light-bending model \citep{miniutti2004}. This model is
also considered for explanation of the soft-excess (Crummy et al. 2006,
Petrucci et al. 2006, Ponti et al. 2006) although being confronted to other
approaches (Gierli\' nski \& Done 2004, Chevallier et al. 2006).

\begin{figure*}
  \centering
  \includegraphics[width=10.5cm]{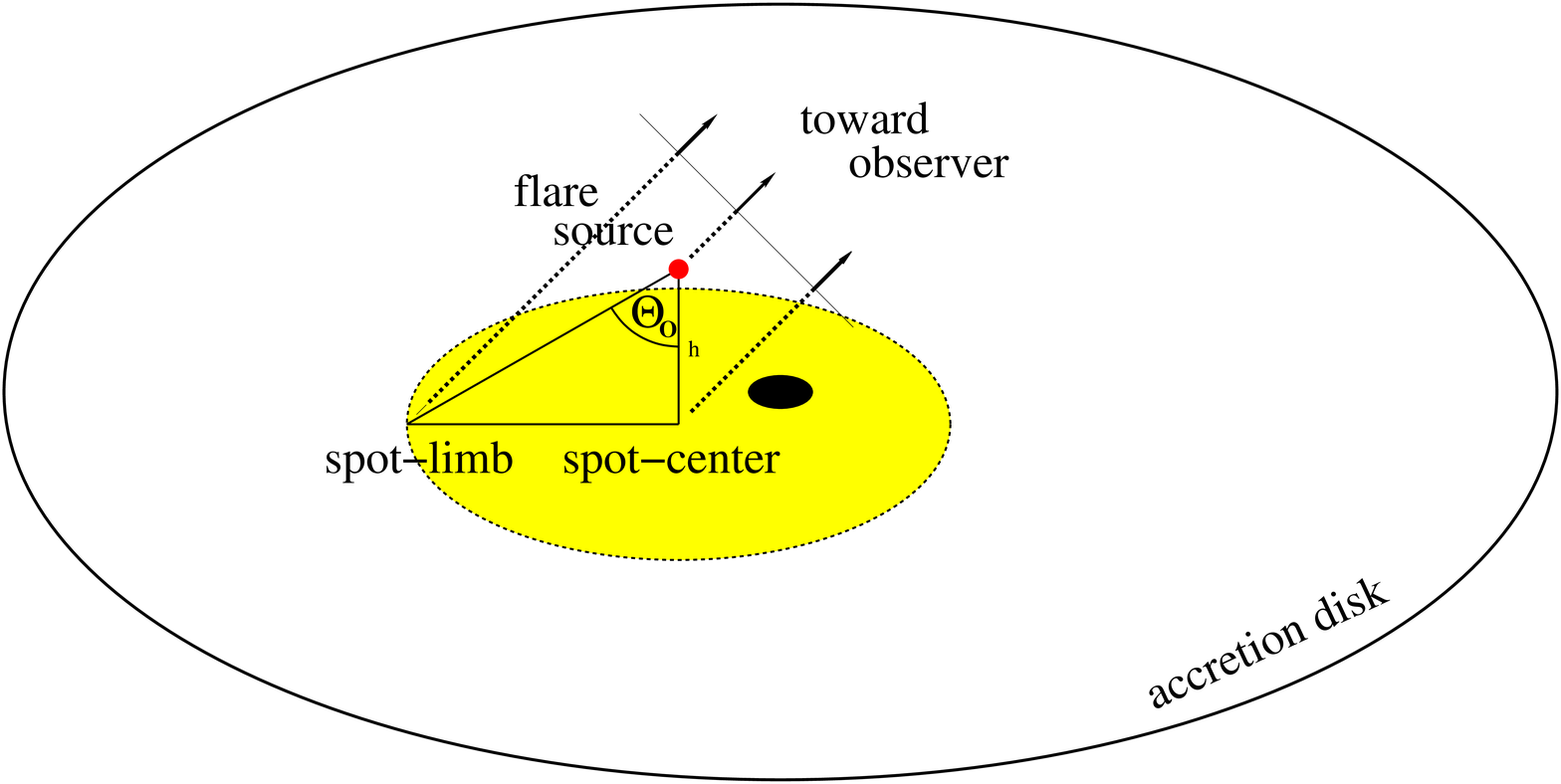}
  \hskip 0.2cm
  \includegraphics[width=7cm]{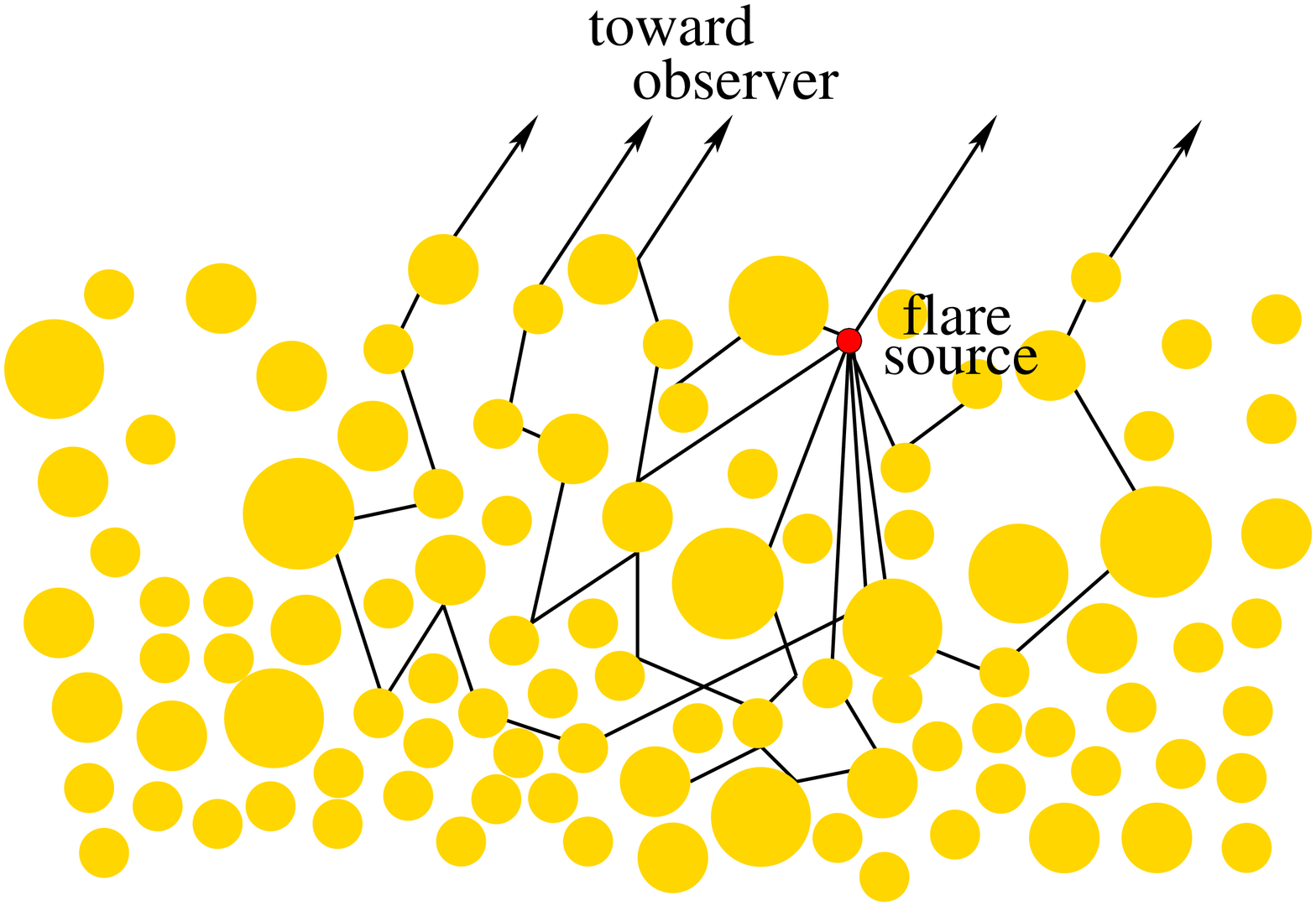}
  \vskip 0.2 cm
  \caption{Geometry of a flare. Left: a compact hard X-ray source above a
  reprocessing accretion disk. Right: a compact source illuminating a clumpy
  accretion flow. The flare is partly embedded in the flow without being
  completely hidden.\label{fig:flare-geom}}
\end{figure*}

We illustrate the simplified geometry of a flare scenario in
Fig.~\ref{fig:flare-geom} (left). The radiation is assumed to be beamed or
bent toward the accretion disk and released within a cone of a certain
half-opening angle $\theta_0$. A larger $\theta_0$ leads to a wider range of
light travel times between the flare source and the disk, and hence the
broadening of $I_{\rm{}r}$ increases. By varying $b$ in our model we thus
change the physical picture assuming a different angular release of the primary
radiation.

The geometry of the light-bending model assumes the flare source to be close
to the disk axis and at varying heights above the disk surface. This lamppost
geometry is justified by the idea of shocks during the relativistic jet
formation (Henri \& Pelletier 1991, Martocchia et al. 2000, 2002). It is
interesting to note that during the 95 ksec observation the detected line
possibly became narrower with time ($\sigma = 0.39 \pm 0.20$ for the detection
3000~sec after the flare peak, and $\sigma = 0.19 \pm 0.12$ for the detection
4000~sec after the flare peak, Ponti et al. 2004), as if coming from more
distant parts of the disk.

The delay model does not constrain the radial position of the flare but it
infers that the flare source is elevated rather high above the disk. Assuming
that the flare is triggered by magnetic reconnection, the height of the flare
source above the disk surface should be equal to the disk`s pressure scale
height. This is implied by the equipartition between thermal and magnetic
pressure as the flux tubes rise into the corona by buoyancy. An analytical fit
of the disk`s pressure scale height was given by Eqn.~(9) in
\citet{czerny2004}. With $M = 3.3 \times 10^6$ M$_\odot$ and assuming a
high-state accretion rate of 0.3 times the Eddington rate \citep{uttley2002},
we obtain pressure scale heights lower than 4.5 $R_{\rm g}$ for radii $R$ with
$6 \, R_{\rm g} < R < 600 \, R_{\rm g}$. From our delay model we infer an
intrinsic time delay $\delta$, which corresponds to an elevation of the flare
source of 60~$R_{\rm g}$. Far from the axis, it is difficult to imagine a
mechanism transporting the flux tubes so high up. Therefore, a lamppost
geometry where the source is driven upwards by a central outflow would seem
more likely. However, as we discuss below there is another difficulty with the
model in which the primary source occurs at such a large height.

\Citet{ballantyne2001b} and \citet{ballantyne2003} invoke two reflectors to
explain the shape of the iron K$\alpha$ line in MCG-6-30-15. They presume
highly ionized reflection close to the black hole and a distant colder
reflector at $\sim 70 R_{\rm g}$. In principle, this cold reflection may also
form if the flare is elevated high above the disk surface as proposed in the
lamp post model. The reflection is weakly ionized or neutral if the ionization
parameter $\xi = L/(4 \pi d^2 n) < 10$. Herein $d$ is the distance of the
flare from the disk surface with density $n$. With $L = 2.6 \times
10^{42}$~erg~s$^{-1}$ and the local density in the disk atmosphere $n =
10^{14}\,$cm$^{-3}$ we obtain the condition for an almost neutral reflection
as $d > 1.5 \times 10^{14}$~cm, which corresponds to a light travel time $>
1$~hour. This is more than the measured delay time during the observation and
makes the scenario of a highly-elevated flare unlikely.

\subsection{The weakly ionized reflector}

The question arises where a weakly ionized/neutral reflector can be located if
the flare is not allowed to be far away from the reprocessor. A possible
answer is a model of cold, clumpy reflectors: the existence of magnetically
confined, cold clouds in the accretion flow was suggested by  Guilbert \& Rees
(1988), and explored further by Celotti, Fabian \& Rees (1992) and Kuncic,
Celotti \& Rees (1997). The interest in this model was recently revived by the
detection of narrow redshifted absorption lines which suggest an inflow over a
narrow range of radii (NGC~3516, Nandra et al. 1999; Mrk~509, Dadina et
al. 2005; E~1821+643, Yaqoob \& Setlemitsos 2005; Q0056-363, Matt et al. 2005;
PG~12111+143, Reeves et al. 2005; Mrk~335, Longinotti et al. 2006). The
magnetically confined clumps were also recently discussed in the context of
the very high state in galactic sources (see e.g. Yuan et al. 2006 and
references therein). In general, partial covering remains an interesting
alternative to the flat disk reflection for a number of Narrow Line  Seyfert 1
galaxies (Mrk~335, Longinotti et al. 2006; 1H~0707-495, Fabian et al. 2002b). 

An illustration of a clumpy reflector model is shown in
Fig.~\ref{fig:flare-geom} (right). We include some light trajectories between
the flare source, the various reprocessing clouds, and the distant
observer. The broadening effect on the reflection signal depends on the exact
geometrical arrangement. In general, a range of light travel times around an
average value is expected. Very low values of our model parameter $b$
correspond to a geometrical arrangement, where the reprocessing clouds are
distributed symmetrically with respect to the flare source and to the
observer. By increasing $b$ our setup drifts away from such a symmetry.

During the flare the iron K$\alpha$ line complex was not detected. A strong
enhancement of the line emission was seen only $\sim 3000 $~sec after the flare
(Ponti et al. 2004). On the other hand, considerable spectral variability was
seen during the flare. It is therefore possible that while the flare went off
the line formed in the clumpy medium and was strongly broadened, as suggested
by the parameter $b$ we obtain from the delay modeling. Then, the line could
not be identified in the data. The line signature became apparent only later
when the flare radiation was subsequently reprocessed by more distant and
larger parts of the disk.

Our delay model requires a significantly enhanced reflection which indicates a
large covering factor of the flare source with reprocessing clouds. This is
consistent with the (generally) large value of the equivalent width of the iron
line in MCG-6-30-15, particularly during the low flux periods (Fabian et
al. 2002a). Malzac et al. (2006) pointed out that strong reprocessing can be
realized by reflection in a clumpy medium, which presents an alternative to
the classical light-bending models, where the flare source is elevated above
the disk.

\subsection{Conclusion for a global picture}

The observed flare is an exceptional event. Our model does not require the
magnetically confined, cold clouds to exist everywhere in the accretion
flow. We rather suggest that MCG-6-30-15 exhibits both, ionized reflection
from elements of an accretion disk and colder reflection from confined
clouds. The particular flare event we discuss here possibly occurred in a
relatively distant region dominated by magnetic confinement and weak
ionization, which correctly reproduces the observed time delays. Eventually,
the magnetic confinement is correlated with the existence of the reconnecting
flux tubes producing the flare.

Evidences of occasional single flares occurring at distances of the order of
$\sim 10 \, R_{\rm g}$ -- $100 \, R_{\rm g}$ from the black hole are given by
the signature of their orbital motion (e.g. Iwasawa et al. 2004, Turner et
al. 2006). Closer in, the continuous disk may recover again due to
condensation, as suggested by Liu et al. (2006) and Meyer et
al. (2006). Therefore, our model is not in contradiction with general models
of various phenomena in this source based on the processes taking place in the
innermost region.

The qualitative results we obtain suggest to continue this line of research:
to perform more sophisticated modeling of time delays and to look for suitable
single flares in X-ray lightcurves of this object and of other AGN. It would
be particularly important to see whether the departure of the time delay --
energy bin relation from a straight line, well reproduced in our model, is a
characteristic property of the phenomenon.

\section*{Acknowledgments}
We thank Martine Mouchet and Michal Dov{\v c}iak for their comments on an
early draft of this paper and we are grateful to Agata R{\'o}{\.z}a{\'n}ska
and Anne-Marie Dumont for their help with computing the vertical disk profile
and conducting the radiative transfer. GP thanks Patricia Ar\'evalo and
Matteo Genghini for their help. Finally, we are grateful to the anonymous
referee for detailed and constructive comments. 

This work was supported by the grants 1P03D00829, and PBZ-KBN-054/P03/2001 of
the Polish State Committee for Scientific Research (BC), by the Laboratoire
Europe\' en Associ\' e Astrophysique Pologne--France (BC and RWG), and by the
Hans-B\"ockler-Stiftung (RWG). VK and RWG acknowledge support through the
grants PECS 98040, IAA\,300030510, GAUK 299/2004, and the Center for
Theoretical Astrophysics in Prague.

\end{document}